\documentstyle[12pt,epsfig]{article}

\begin{document}


%
%

\newcommand{\sslash}[1]{\rlap/ #1}
\newcommand{\trh}{T_{\rm rh}}
\newcommand{\teff}{T_{\rm eff}}
\newcommand{\MP}{M_{\rm P}}
\newcommand{\delcp}{\delta_{_{\rm CP}}}
\newcommand{\nb}{n_{\rm B}}
\newcommand{\nl}{n_{\rm L}}
\newcommand{\mh}{m_{_{\rm H}}}
\newcommand{\mw}{m_{_{\rm W}}}
\newcommand{\alphaw}{\alpha_{_{\rm W}}}
\newcommand{\ncs}{N_{_{\rm CS}}}
\newcommand{\half}{{1\over2}}
\newcommand{\be}{\begin{equation}}
\newcommand{\ee}{\end{equation}}
\newcommand{\bea}{\begin{eqnarray}}
\newcommand{\eea}{\end{eqnarray}}


\begin{titlepage}
\begin{flushright}
CERN-TH/2001-254\\
hep-ph/0109230
\end{flushright}

\begin{center}
\vspace{1cm}

{\Large \bf Particle production from symmetry breaking \\[3mm] 
after inflation and leptogenesis}

\vspace{0.8cm}

{ \bf Juan Garc{\'\i}a-Bellido \ and \ Ester Ruiz Morales}

\vspace{0.2cm}

{\it Theory Division, CERN, CH-1211 Gen\`eve 23, Switzerland}

\end{center}
\vspace{1cm}

\begin{abstract}
  
  Recent studies suggest that the process of symmetry breaking after
  inflation typically occurs very fast, within a single oscillation of
  the symmetry-breaking field, due to the spinodal growth of its
  long-wave modes, otherwise known as `tachyonic preheating'. We show
  how this sudden transition from the false to the true vacuum can
  induce a significant production of particles, bosons and fermions,
  coupled to the symmetry-breaking field. We find that this new
  mechanism of particle production in the early Universe may have
  interesting consequences for the origin of supermassive dark matter
  and the generation of the observed baryon asymmetry through
  leptogenesis.
 
\end{abstract}

\vspace{1cm}

\vspace{0.5cm}



\end{titlepage}


\renewcommand{\thefootnote}{\arabic{footnote}}
\setcounter{footnote}{0}

\section{Introduction}

Spontaneous symmetry breaking (SSB) is one of the basic ingredients of
modern theories of elementary particles. It is usually assumed that
SSB in Grand Unified and Electroweak theories took place in the early
Universe through a thermal phase transition. However, it is also
possible that some of these symmetries were broken at the end of a
period of inflation~\cite{GBKS}, when the Universe had zero
temperature and the negative mass term for the Higgs field appeared
suddenly, i.e. in a time scale much shorter than the time required
for SSB to occur. In this case, as was recently shown in
Refs.~\cite{GBKLT}, the process of symmetry breaking is extremely
fast.  The exponential growth of the Higgs quantum fluctuations is so
efficient that SSB is typically completed within a single oscillation,
while the field rolls down towards the minimum of its effective
potential. This process, known as tachyonic preheating, leads to
an almost instant conversion of the initial vacuum energy into
classical waves of the scalar fields, in contrast with the process of
`parametric preheating', in which the inflaton field performs many
oscillations before reheating the Universe~\cite{KLS}. 

In this letter we describe how this sudden transition from the false to
the true vacuum can induce the non-adiabatic production of particles
coupled to the Higgs. We also studied the consequences that this new
process may have on the generation of the dark matter and the baryon
asymmetry via leptogenesis. The phenomenon of particle production 
from symmetry breaking is analogous to the Schwinger
mechanism~\cite{Schwinger}, where the role of the external electric
field pulse is played here by the time-dependent expectation
value of the Higgs field. It is also similar to the well known process 
of particle production by a time-dependent gravitational 
background~\cite{Parker}, responsible for the observed anisotropies of 
the microwave background, as well as for Hawking radiation~\cite{Hawking}. 
The difference with respect to the standard mechanism of gravitational
particle production in an expanding Universe is that, in our case,
quantum excitations of fields coupled to the Higgs are produced
due to the sharp change in their masses during the process
of symmetry breaking, instead of due to the quick growth of the
scale factor.  

We will consider here a simplified model of SSB in which the Higgs
instantly acquires a negative mass-squared term~\cite{GBKLT}.
This `quench' approximation corresponds to the limiting case of a
hybrid inflation model~\cite{hybrid} satisfying the so-called
`waterfall' condition. We therefore assume that the complex symmetry
breaking field $\phi$ starts in the false vacuum at the top of its
potential $V(\phi)=\lambda(|\phi|^2-v^2)^2/4$, with zero mean,
$\langle\phi\rangle= 0$, and initial conditions given by vacuum
quantum fluctuations. We will study the production
of other fields, scalars $\chi$ and fermions $\psi$, coupled to the
Higgs with the usual scalar $g^2|\phi|^2\chi^2$ and Yukawa
$\,h\phi\bar\psi\psi\,$ interactions.  As we will show later, the
backreaction of the produced $\chi$ and $\psi$ modes on the evolution
of the Higgs expectation value (vev) is negligible for the small $g$
and $h$ couplings we are considering, so that we can solve first the
process of symmetry breaking, and then take the resulting evolution of
the Higgs as a background field that induces particle production.

\section{Symmetry Breaking}

The dynamics of symmetry breaking has been studied in detail in
Refs.~\cite{GBKLT}.  Here we only summarise the main results needed
for our analysis.  At the initial stages of SSB, when re-scattering
effects are still unimportant, the Higgs modes follow the linear
equation $\ddot\phi_k+(k^2 - m^2)\,\phi_k = 0$, where
$m^2=\lambda\,v^2$.  With vacuum initial conditions,
$\phi_k(0)=1/\sqrt{2k},\ \dot\phi_k(0)=-ik\,\phi_k(0)$, all
long-wavelength modes within the horizon ($H<k<m$) grow exponentially,
$\phi_k(t) = \phi_k(0) \exp(t\sqrt{m^2-k^2})$, while modes with $k>m$
oscillate with constant amplitude. The exponential growth continues
until the long-wave modes reach a value for which the effective Higgs
mass becomes positive, i.e. when $\langle|\phi|^2\rangle \geq
v^2/\sqrt3$, and the symmetry is broken soon after.

\begin{figure}[t]
\vspace{-.5cm}
\begin{center}
\includegraphics[width=10cm,height=14cm,angle=-90]{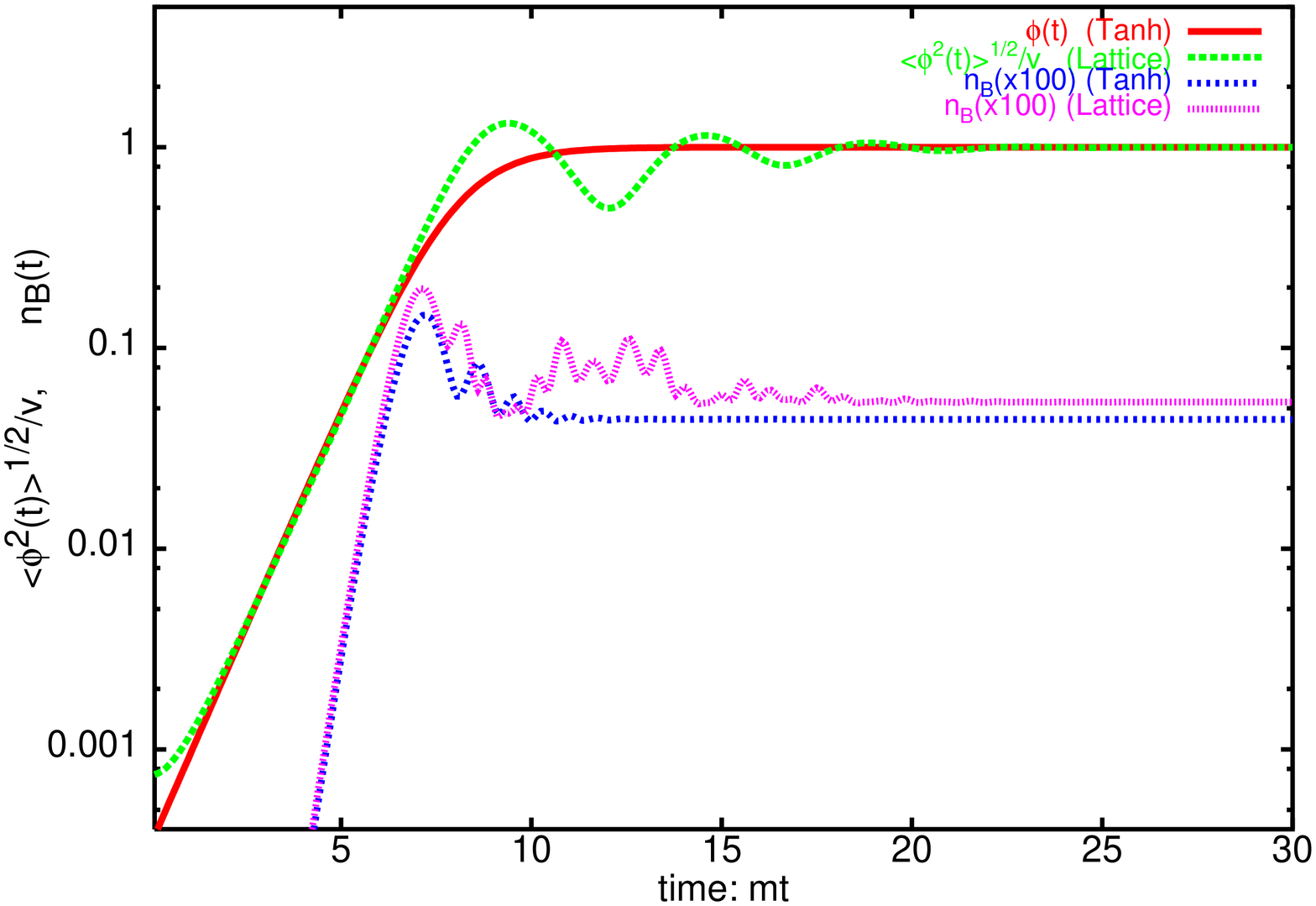}
\end{center}
\caption{}\label{fig1}
\end{figure}

We have chosen $n_k + \half = |\phi^*_k(t)\dot\phi_k(t)|$ as a proper
definition for the occupation numbers of the Higgs tachyonic modes
\cite{jan}.
This expression does not require an {\it a priori} definition of a
mode frequency $\omega_k$, and matches smoothly the one used in
\cite{GBKLT} for positive frequencies. For the growing modes, we have
occupation numbers
\be\label{phink}
n_k + \half = \half\Big|e^{t\sqrt{m^2-k^2}}\Big|^2 \approx 
\half \,e^{2mt}\, e^{-{k^2 \over 2k_*^2}}\,,
\ee 
that become exponentially large very quickly for the long wavelength modes, 
and drop abruptly for $k>k_* = m\,(2mt)^{-1/2}$, which gives a natural 
ultraviolet cutoff for the problem. Note that, while the dynamics conserves
$\langle\phi\rangle=0$, the Higgs dispersion also grows exponentially
\be
\langle|\phi|^2\rangle = {m^2\over8\pi^2}\int_0^m{dk^2\over m^2}\,
e^{2t\sqrt{m^2-k^2}} = {m^2\over16\pi^2}\,{e^{2mt}(2mt-1)+1\over 
m^2t^2} \sim {m^2\over16\pi^2}\,e^{2mt}\,, \label{RMSE}
\ee
where this expression has been regularised, as described below. The
time it takes for the system to break the symmetry, i.e. when
$\langle|\phi|^2(t_*)\rangle\simeq v^2$, can be estimated as 
\be\label{t*}
m\,t_* \simeq \half\log\Big({32\pi^2\over\lambda}\Big)\,. 
\ee
This time depends only logarithmically on the Higgs self-coupling
constant $\lambda$. For typical values, $\lambda=10^{-3}$ and
$v=10^{-4}\MP$, the symmetry is broken within $m t_* \sim 6$,
and the typical cut-off frequency becomes $k_* \sim m/3$. This means
that, by that time, the occupation numbers of modes with $k<k_*$ is
exponentially large,
\be
n_k(t_*) \approx \half e^{2mt_*} \simeq {16\pi^2\over\lambda} \sim 
2\times 10^5\,.  
\ee
These large occupation numbers allow us to treat these modes as
semiclassical waves and match the solutions of the linear equations
with the fully non-linear numerical lattice simulations~\cite{KTPR}.
The non-linear dynamics is studied by solving the real time evolution
equations of classical fields, using a modified version of the lattice
simulation program LATTICEEASY of Felder and Tkachev~\cite{FT}. We
start with initial fluctuations described by a Gaussian random field
with zero mean, $\langle\phi_k\rangle=0$, and regularised dispersion
\be
\langle|\phi_k^2(t)|\rangle_{\rm reg} \equiv 
\langle|\phi_k^2(t)|\rangle - {1\over2\omega_k}\,.
\ee
This prescription amounts to
substituting quantum averages by ensemble averages and ensures that
the physical masses and energies are not ultraviolet divergent.  In
Fig.~\ref{fig1}, we show the result of the full non-linear evolution
of the Higgs vev, and compare it with the approximate expression
\begin{equation}\label{phi}
\phi(t)\equiv \langle|\phi|^2(t)\rangle^{1/2}={v\over2}
\Big(1+\tanh{m(t-t_*)\over2}\Big)\,,
\end{equation}
which ignores the strongly damped oscillations after symmetry
breaking~\cite{GBKLT}. We have checked that these low-amplitude
oscillations do not contribute to the non-adiabatic production of
particles; i.e. parametric preheating is inefficient after symmetry
breaking, a result anticipated in Ref.~\cite{GBL} for the case of
hybrid inflation. Note also that we are using the Higgs vev averaged
over the lattice as a homogeneous background field, while it is
actually a sum of tachyonic modes with different frequencies. However,
in analogy with the generation of anisotropies during inflation, what
drives the growth of long-wave modes $k<k_*$ of particles coupled to the
Higgs is the coarse-grained Higgs vev over scales $\sim k^{-1}$, which
is practically the same as the average over the whole Hubble volume,
thanks to the fact that all Higgs modes with $H^{-1}<k<k_*$ grow
essentially at the same speed.

\section{Particle Production}

We can now calculate the production of bosons and fermions coupled to
the Higgs using the formalism of quantum fields in strong
backgrounds~\cite{GMM,BD}. The mode equations 
in terms of rescaled fields, $X_k(t)=a^{3/2}\chi_k$ and $\Psi(t)=
a^{3/2}\psi$, are
\begin{eqnarray}\label{boseq}
\partial_t^2 X_k + \Big(k^2 + m^2_{\rm B}(t)a^2(t)\Big) X_k &=& 0\,,\\
\Big(i\gamma^\mu\partial_\mu - m_{\rm F}(t)a(t)\Big) \Psi &=& 0\,,
\label{fereq}
\end{eqnarray}
where both the mass $m(t)$ and the scale factor $a(t)$ depend on time.
In hybrid inflation models for which the quench approximation is valid,
the rate of expansion is typically much smaller than the masses involved, 
and we can take the scale factor to be constant ($a=1$) during SSB. 
We will only consider here the non-adiabatic production of particles due 
to the change of vacuum as it induces a sudden change in the inertia (masses)
of bosons $m^2_{\rm B}(t)= g^2\langle|\phi|^2\rangle$ and fermions 
$m_{\rm F}(t)=h\langle|\phi|^2\rangle^{1/2}$, through the Higgs mechanism.

We have solved the mode equations (\ref{boseq}) and (\ref{fereq}) both
numerically, using the lattice Higgs vev evolution, and within the
approximation (\ref{phi}), for which one has analytical solutions in
terms of Hypergeometric functions~\cite{GMM}. The number density of
created particles, as seen by a future observer in the true vacuum, is
then given by 
\be
n_X = {g_s\over2\pi^2a^3}\,\int\,dk\,k^2\,|\beta_k|^2\,,
\ee
where $X$ denotes either bosons or fermions; $\beta_k$ are the
Bogoliubov coefficients that relate the {\it in} ($t\to-\infty$) and
{\it out} ($t\to+\infty$) mode functions; $k$ is the comoving
momentum, and we have summed over spin indices ($g_s=1$ for scalars, 2
for spinors). In the case of charged fields, $n_X$ gives the number of
particles, equal to that of antiparticles (i.e.  $g_s=2$ for complex
scalars, 2 for Majorana fermions, 4 for Dirac fermions). 

Let us consider first the production of bosons. The mode functions
$X_k(t)$ of the scalar field are solutions of the oscillator equation
(\ref{boseq}) with time-dependent frequency $\omega_k^2=k^2+m_B^2$,
and initial vacuum conditions, $X_k(0)=1/\sqrt{2\omega_k}$ and
$X'_k(0)=-i \omega_k X_k$. In this case, the Bogoliubov coefficient at
$t\to\infty$ can be written as $n_k^{\rm B}=|\beta_k|^2=(|X'_k|^2+
\omega_k^2|X_k|^2-\omega_k) /2\,\omega_k$. The exact solution of the 
bosonic equation,
\be
X''_k + \left(\omega_{-}^2(k) + {\alpha^2\over4}(1+\tanh\,x)^2\right)\,
X_k = 0\,,
\ee
with $x=m(t-t_*)/2$, can be written in terms of Hypergeometric
functions as~\cite{GMM}
\be\label{solB}
X_k=C_1(k)\,e^{ikt}(1+e^{mt})^\tau\,F(a,b,c;-e^{mt})+
C_2(k)\,e^{-ikt}(1+e^{mt})^\tau\,F(a^*,b^*,c^*;-e^{mt})\,,
\ee
where $\tau$ is a solution of the equation $\tau(\tau-1)+\alpha^2=0$,
and
\be
a = \tau + i(\omega_{-}-\omega_{+})\,,\hspace{1cm} 
b = \tau + i(\omega_{-}+\omega_{+})\,,\hspace{1cm}
c = 1 + 2i\,\omega_{-}\,,
\ee
where $\omega_\pm$ are the {\it in/out} asymptotic frequencies,
$\omega_{-}(k)=k$ and $\omega_{+}(k)=\sqrt{k^2+\alpha^2m^2}$, and
$\alpha\equiv g/\sqrt\lambda$.
These solutions match the initial conditions at the end of
inflation, and can be analytically continued to $t\to\infty$,
in order to compute the boson occupation number $n_k^{\rm B}=
|\beta_k|^2$,
\begin{equation}\label{nbos}
n_k^{\rm B}(\alpha) =
{\cosh[\pi\sqrt{4\alpha^2-1}] + \cosh[2\pi(\omega_{+}-\omega_{-})/m]
\over\sinh[2\pi\omega_{-}/m]\,\sinh[2\pi\omega_{+}/m]}\,.
\end{equation}

A similar analysis can be done in the case of fermions, where the
first-order Dirac equation (\ref{fereq}) for the two spinor components
can be written as an oscillator equation with complex
frequency, $X''_k+(k^2+m_F^2-im_F')\,X_k = 0$, see Ref.~\cite{GMM},
\be
X''_k + \left(\omega_{-}^2(k) + {\alpha^2\over4}(1+\tanh\,x)^2
-{i\alpha\over4}(1+\tanh^2\,x)\right)\,X_k = 0\,.
\ee
Given the initial vacuum conditions, $X_k(0)=(1-m_F/\omega_k)^{1/2}$
and $X'_k(0)=-i \omega_k X_k$, we can write the solution as in
Eq.~(\ref{solB}), but with $\tau = -i\alpha$. The fermionic occupation
number can then be written as $n_k^{\rm F}=|\beta_k|^2=
(\omega_k-m_F-{\rm Im}X_k{X_k'}^*)/2\,\omega_k$. Substituting the
$t\to\infty$ solutions of the fermionic equations we find
\begin{equation}\label{nfer}
n_k^{\rm F}(\alpha) =
{\cosh[2\pi\alpha] - \cosh[2\pi(\omega_{+}-\omega_{-})/m]\over
2\sinh[2\pi\omega_{-}/m]\,\sinh[2\pi\omega_{+}/m]}\,,
\end{equation}
with $\omega_{-}$ and $\omega_{+}$ the same as for bosons but with
$\alpha\equiv h/\sqrt\lambda$.

The occupation numbers obtained numerically from Eqs. (\ref{boseq})
and (\ref{fereq}), using the full non-linear lattice solution for the
Higgs vev, agree very well with the analytical formulae (\ref{nbos})
and (\ref{nfer}), see Fig.~2,
except at large momenta, where there is a small contribution from
resonant particle production due to the strongly damped Higgs
oscillations.

We will now calculate the ratio of energy densities of the particles
created at the end of symmetry breaking to the initial false vacuum
energy density
\be
{\rho_X\over\rho_0} = {2g_s\lambda\over\pi^2}\int\,d\kappa\,
\kappa^2\,n_\kappa^X(\alpha)\,\omega_+(\kappa,\alpha),
\ee
where $\kappa=k/m$ and $\rho_0 = m^4/4\lambda$. We can obtain a fit 
to the final energy density of bosons and fermions as 
\bea
\rho_{\rm B}/\rho_0 & \simeq &
2\times10^{-3}\,g_s\lambda\, f(\alpha,1.3)\,, \label{rhob}\\
\rho_{\rm F}/\rho_0 & \simeq & 
1.5\times10^{-3}\, g_s\lambda\,f(\alpha,0.8)\,,
\label{rhof}
\eea
where the fitting function is $f(\alpha,\gamma) \equiv 
\sqrt{\alpha^2+\gamma^2}-\gamma$, and the coefficient
$\alpha=g/\sqrt{\lambda}\,(h/\sqrt{\lambda})$ for bosons (fermions), 
has been fitted in the range $\alpha\in[0.01,10]$.

We can also write the number densities for bosons and fermions,
\bea
n_{\rm B}(\alpha) & \simeq &
1\times10^{-3}g_s\,m^3\,f(\alpha,1.3)/\alpha\,, \label{nub}\\
n_{\rm F}(\alpha) & \simeq & 3.6\times10^{-4}g_s\,m^3\,f(\alpha,0.8)/\alpha\,. 
\label{nuf}
\eea
From these expression we see that the production of bosons and
fermions from symmetry breaking is more efficient when the Higgs mass
and thus $\lambda$ is large, driving a very steep growth of the vev
towards the true vacuum. In the case of very large couplings $g^2,
h^2 \gg \lambda$, the produced particles are non-relativistic,
$\omega_k\approx m_X$, and their energy density is given by $\rho_{\rm
X} = m_{\rm X} n_{\rm X}$. Although the energy density increases with
the mass of the particles produced, their number density saturates for
large $\alpha$ to a value proportional to $m^3$, and thus the number
of particles produced cannot be made arbitrarily large by increasing
$g$ and $h$ couplings. Furthermore, the physical momenta redshift as
$k_{\rm phys} = k/a$ with the expansion of the Universe, and thus the 
number density of both relativistic
and non-relativistic bosons or fermions decays as $n_X \sim a^{-3}$.

Note that, unless the couplings are unnaturally large, the fractional
energy density in bosons and fermions is always small, so we do not
expect an important backreaction on the evolution of the Higgs
condensate as the symmetry is broken. Moreover, contrary to the case
of Ref.~\cite{FK}, in which particles are produced long after symmetry
breaking from non-linear rescattering, our mechanism of particle
production from symmetry breaking gives an upper limit to the
occupation numbers of bosons produced in the range $H<k<m$, even for
arbitrarily large coupling $g$, which is of order $n_k \leq 1$.  This
prevents us from using LATTICEEASY to compute their energy density and
backreaction.

We would like to explore now the cosmological consequences that this
production of particles may have for the evolution of the Universe. 

\section{Leptogenesis}

Leptogenesis from preheating was first proposed in Ref.~\cite{GPRT} in
the context of chaotic inflation, where the right-handed (RH)
neutrinos were produced via parametric resonance through their
coupling to the inflaton.\footnote{Fermion production during
  preheating via parametric resonance was also studied in
  Ref.~\cite{paraferm}.}  Here we propose a novel scenario in which a
large population of out-of-equilibrium RH neutrinos is produced in the
process of symmetry breaking after hybrid inflation, via their
coupling to the SSB field.

Let us suppose that the symmetry-breaking field responsible for the
end of hybrid inflation, and undergoing tachyonic preheating, is in
fact the Higgs field of a grand unified theory (GUT) with
($B-L$)-violating interactions, and that the same Higgs produces a
significant population of right-handed Majorana neutrinos via the
non-perturbative mechanism described above.  These massive neutrinos
decay (out of equilibrium) into leptons and standard model (SM)
Higg\-ses, generating a leptonic asymmetry~\cite{FY} that later gets
converted into the baryon asymmetry of the Universe via sphaleron
transitions~\cite{KS}.

The Lagrangian terms relevant for leptogenesis are
\be
{\cal L} = - \half\,h\,\bar{N}^c_R\,\phi\,N_R - h_\nu\,\bar{N}\,H\,l_L\,,
\ee
where the first term determines the masses of RH neutrinos through the
GUT Higgs mechanism, $M_R = h\,\langle\phi\rangle = h\,v$, while the
second term determines the lepton-violating decays of the RH
neutrinos into SM Higgses $H$ and left-handed leptons $l_L$.  Although
both Yukawas $h$ and $h_\nu$ are actually 3x3 matrices, we will not
consider here the complications of 3-generation neutrino masses and
mixings, but will assume as usual~\cite{BP} that only the lightest RH
neutrino $N_1$, with mass $M_1$, is responsible for leptogenesis.  For
analogous reasons, the CP-violating asymmetry $\varepsilon$ in the
decay of the RH neutrino, which also depends on the structure of
neutrino masses and mixing, will be taken as a free parameter.

Assuming that the masses of the SM light neutrinos are
generated by the see-saw mechanism~\cite{seesaw},
$m_{\nu_i}=(m_Dm_D^\dagger)_{ii}/M_i$, the decay rate of the lightest
RH neutrino can be written in terms of $m_{\nu_1}$ and $M_1$ as
\begin{equation}
\Gamma_{N_1} = {G_F\over2\sqrt2\,\pi}\,m_{\nu_1}\,M_1^2
= 4.12\times10^6\,h^2\,
\left({v\over10^{15}\,{\rm GeV}}\right)^2
\left({m_{\nu_1}\over10^{-9}\,{\rm eV}}\right)\,{\rm GeV}\,.
\end{equation}

Under these assumptions, the model has 5 parameters: the vev of the
GUT Higgs $v$, the Higgs self-coupling $\lambda$, the Yukawa coupling
of the RH neutrino $h$, the mass of the light neutrino $m_{\nu_1}$,
and the final reheating temperature $\trh$. We will now study the
constraints that will determine the allowed range of parameters for
a successful model of leptogenesis.

The first requirement is that the RH neutrinos produced
at symmetry breaking decay out of equilibrium. To ensure this we
impose two conditions: i) that the Higgses produced via tachyonic
preheating do not reach thermal equilibrium before the RH neutrinos
decay; and ii) that the Higgs-neutrino interaction rate is much smaller
than the rate of expansion of the Universe, so that even if the
Higgses thermalise, the RH neutrinos remain out of equilibrium.

In order to satisfy condition i), we require that the Higgs 
self-interaction rate at the time of symmetry breaking is smaller
than the rate of expansion, $\Gamma_{\rm self} < H_{\rm SB} = 
(2\pi/3)^{1/2}\,m\,v/\MP$, where
\be
\Gamma_{\rm self} = \int\,{d^3k\over(2\pi)^3}\,\sigma_{\rm self}\,
n_k^\phi(t_*)\,{k\over m} = {2\lambda m\over\pi\ln(32\pi^2/\lambda)}\,,
\ee
with $\sigma_{\rm self}={\lambda^2\over8\pi k_*^2}$ and $n_k^\phi(t_*)=
{16\pi^2\over\lambda}\,e^{-k^2/2k_*^2}$. Note that, once ensured at
symmetry breaking, this condition is always satisfied since $n_\phi(t)
\sim a^{-3}$ while $H \sim t^{-1}$. This condition constrains the Higgs
self-coupling to be
\be\label{lbound}
{\lambda\over\ln(32\pi^2/\lambda)} < \Big({\pi^3\over6}\Big)^{1/2}\,
{v\over\MP}\,.
\ee
Once chosen a GUT scale $v$, this constraint gives an upper bound on 
$\lambda$, and thus on the maximum number of RH neutrinos produced in 
this model. We will take $v = 10^{15}$ GeV, and $\lambda = 10^{-3}$,
which satisfies the bound.

The second constraint ii) requires $\Gamma_{\rm int} < H_{\rm SB}$, where
\be
\Gamma_{\rm int} = \int\,{d^3k\over(2\pi)^3}\,\sigma_{\rm int}\,
n_k^\phi(1-n_k^F)\,{k\over m} = {h^2 m\over\pi^2\ln^2(32\pi^2/\lambda)}\,,
\ee
with the Higgs-$N_1$ cross-section given by $\sigma_{\rm int}={h^4\over
16\pi^2M_1^2}$, and where we have neglected
the fermion contribution $n_k^F\ll1$. This gives
\be\label{hbound}
{h\over\ln(32\pi^2/\lambda)} < \Big({2\pi^5\over3}\Big)^{1/4}\,
\Big({v\over\MP}\Big)^{1/2}\,.
\ee
For the chosen parameters, we can satisfy this constraint as long as
$h<0.5$. We will choose $h=0.01$, i.e. $M_1 = 10^{13}$ GeV. With these
values of the parameters, the number density (\ref{nuf}) and
fractional energy density (\ref{rhof}) of RH neutrinos at symmetry
breaking becomes $n_{N_1} \simeq 1.4\times10^{-4}\,m^3$ and
$\rho_{N_1}/\rho_0 \simeq 7.4\times10^{-6}$.

We will now study the constraints associated with the time of decay of
the RH neutrinos, $t_1 = \Gamma_{N_1}^{-1}$. Since $v$, $\lambda$ and
$h$ have been determined by the previous constraints, these new
conditions will give an allowed range for $m_{\nu_1}$. First of all,
the $N_1$ lifetime should be greater than the time of symmetry
breaking, $t_1\geq t_*$, with $t_*$ given in (\ref{t*}). For our
parameters, we should satisfy
\be\label{m1cons}
m_{\nu_1} < 0.5\ {\rm eV}\left({10^{15}\,{\rm GeV}\over
v}\right)\,{\sqrt\lambda\over h^2\ln(32\pi^2/\lambda)}
= 12\ {\rm eV}\,.
\ee
We also have to ensure that the annihilation rate of $N_1$ into
Higgses, $\Gamma_{\rm ann} = n_{N_1}\,\sigma_{\rm ann}$, with
annihilation cross-section $\sigma_{\rm ann}=h^4/16\pi^2M_1^2$, is
smaller than the decay rate, $\Gamma_{\rm ann} < \Gamma_{N_1}$,
otherwise no RH neutrinos would be left to produce the lepton
asymmetry. This imposes the constraint
\be
m_{\nu_1} > 1.1\times10^{-6}\ {\rm eV}\,{\lambda^2\over h}\,
f(h/\sqrt\lambda,0.8)\,\left({10^{15}\,{\rm GeV}\over
v}\right) = 7\times10^{-12}\ {\rm eV}\,, 
\ee
which gives, together with (\ref{m1cons}), a wide range of allowed
$m_{\nu_1}$ values.

Finally, we want our RH neutrinos to decay before the Universe
reheats via the decay of the Higgs, $\Gamma_{N_1} > \Gamma_\phi = 
H(t_{\rm rh})$. This implies a bound on the reheating temperature,
$\trh \sim 0.12\sqrt{\Gamma_\phi\MP}$~\cite{book},
\be
\trh < 2\times 10^{-3}\,M_1\,
\left({m_{\nu_1}\over10^{-9}\,{\rm eV}}\right)^{1/2}\,.
\ee
For $M_1 = 10^{13}$ GeV, and chosing the mass of the light neutrino to
be $m_{\nu_1} = 10^{-9}$ eV, we find that the reheating temperature
should satisfy $\trh < 2\times10^{10}$ GeV, which
is compatible with the bounds coming from gravitino production.

Now we should estimate the effective temperature of the decay products
of the RH neutrinos $\tilde T_1$, assuming that their energy density
at the time of decay $t_1$ is converted into a thermal bath
$\rho_{N_1}(t_1) = {\pi^2\over30} g_* \tilde T_1^4$. The neutrino energy
density at $t_1$ can be calculated from its value at symmetry breaking
as 
\be
\rho_{N_1}(t_1) = \left({ \rho_{N_1}\over \rho_\phi} \right)_{\rm SB}
\left( { H_{\rm SB}\over \Gamma_{N_1}}\right)^{1/2}\rho_\phi(t_1),
\label{rhon1}
\ee 
taking into account that, from $t_{\rm SB}$ until $t_1$, the RH neutrinos 
are non-relativistic and thus their energy density decays like matter, 
$\rho_F \sim a^{-3}$, while the energy density in Higgs particles is
dominated by the gradient and kinetic terms, with a radiation equation
of state, $\rho_\phi \sim a^{-4}$. This gives the extra factor 
$a(t_1)/a(t_{\rm SB})$ in (\ref{rhon1}) coming from the expansion of 
the Universe. Since at $t_1$ the energy density is dominated by the
inflaton, we have $\rho_\phi(t_1)=3\MP^2\Gamma_{N_1}^2/8\pi$.
For our parameters, we find 
\be\label{T1}
\tilde T_1 = 2\times10^{12}\ {\rm GeV}\,h^{3/2}\,\lambda^{5/16}\,
f^{1/4}(h/\sqrt\lambda,0.8)\,\left(v\over{10^{15}\,{\rm GeV}}\right) =
4\times 10^9\ {\rm GeV}\,,
\ee
which does not constitute a problem with gravitino production.

A final concern is that the lepton-number-violating processes among
the particles in this thermal bath are always out of equilibrium, so
that they do not wash out the lepton asymmetry produced,
$\Gamma_{\Delta L=2}= {4\over\pi^3}\tilde T_1^3\,G_{\rm F}^2\, \sum
m_{\nu_i}^2 < H(\tilde T_1)$. We have included here the possible
channels of lepton-number violating interactions mediated by the heavy
RH neutrinos $N_2$ and $N_3$. This bound gives
\be
\tilde T_1 < 3.3\times 10^{13}\ {\rm GeV}\,
\left( {3\times10^{-3}\,{\rm eV}^2\over \sum m_{\nu_i}^2}\right)\,,
\ee
which is easily satisfied, see Eq.~(\ref{T1}), for the range of
light neutrino masses consistent with neutrino 
oscillations~\cite{BGP}.

We may now ask how efficient is this mechanism for producing the
required amount of baryons in the Universe, $\nb/s = (4-9)\times
10^{-11}$, where $s$ is the entropy density, and we use the fact that
this ratio remains constant since reheating.

The baryon asymmetry is produced via sphaleron transitions that
violate $(B+L)$ and convert a lepton asymmetry into a baryon
asymmetry, $\nb=(28/79)\nl$~\cite{KS}. The entropy density at
reheating can be computed in terms of the energy density,
$s=(4/3)\rho_\phi/\trh$, and can be estimated as
\begin{eqnarray}
\left.{\nb\over s}\right|_0 &=& 
\left.{28\over79}\,{\nl\over s}\right|_{\rm rh} = 
\left.{28\over79}\,{3\trh\over4}\,{\nl\over\rho_\phi}\right|_{\rm rh} = 
{28\over79}\,{3\nl\over4n_{N_1}}\,\trh\,
\left.{n_{N_1}\over\rho_\phi}\right|_{\rm SB}
\left({\rho_0\over\rho_{\rm rh}}\right)^{1/4}\nonumber\\
&=& 2.2\times10^{-4}\,\varepsilon\,{\lambda^{5/4}\over h}\,
f(h/\sqrt\lambda,0.8)\,,\label{nbs}
\end{eqnarray}
where we have assumed that the lepton number density is directly
related to the number density of RH neutrinos via $\nl = \varepsilon\,
n_{N_1}$. For the values of the parameters we have chosen, the baryon
asymmetry becomes $\nb/s = 2.3\times10^{-7}\,\varepsilon$, which
requires a relatively large leptonic asymmetry $\varepsilon = 3.4
\times 10^{-4}$, that could be obtained in realistic neutrino models
with large mixing angles. However, significantly smaller values of
$\varepsilon$ can still be accommodated in this scenario since the
values we chose for $h$ and $\lambda$ were very conservative. If we
keep the GUT vev at $10^{15}$ GeV, we can still increase the
production of RH neutrinos by taking the maximum values of the
couplings allowed by the bounds (\ref{lbound}) and (\ref{hbound}),
$h=0.5$ and $\lambda=3\times 10^{-3}$, which gives $\varepsilon \geq 7
\times 10^{-5}$ for agreement with the observed baryon asymmetry of
the Universe. Moreover, the value of $\lambda$ can be chosen greater
than the bound (\ref{lbound}) if we accept that the Higgses thermalise
(before decaying) soon after symmetry breaking, while keeping $h$
small (\ref{hbound}), so that the RH neutrinos remain out of
equilibrium before decaying.

We can also choose a larger vev for the GUT Higgs, e.g. $v=10^{16}$
GeV, thus allowing a larger RH neutrino production. This vev value
would be compatible with all the bounds, and permits a maximum value
for $\lambda=0.02$ and $h=1.15$, which gives $\nb/s = 1.12\times
10^{-5}\,\varepsilon$. This corresponds to values of the leptonic
asymmetry as small as $\varepsilon=7\times10^{-6}$, to be in agreement
with the observed baryon asymmetry of the Universe. Smaller values of
$\varepsilon$ would be difficult to accommodate in our scenario.

We conclude that our scenario can generate a successful leptogenesis
for a wide range of model parameters, although in order to go beyond
our estimates, we would need more information on the structure of
neutrino masses and mixings.

\section{Other relics, dark matter and cosmic rays}

This out-of-equilibrium production of particles during SSB may have
other interesting cosmological consequences. For instance, this
mechanism may be responsible for the non-thermal production of
superheavy dark matter (SDM)~\cite{CKR}. Consider the following
scenario: a GUT scale symmetry breaking at the end of a period of
hybrid inflation, where the SB field may generate via the mechanism
described above, a large population of particles out of equilibrium,
which may either decay into stable relics, or be stable themselves.
Such a heavy relic will eventually dominate the energy density of the
Universe and constitute today the cold dark matter component,
$\Omega_{\rm cdm}h^2 = 0.14\pm0.04$, $95\%$ c.l.  \cite{MS}, where
$h=0.72\pm0.07$ is the present Hubble rate in units of 100 km/s/Mpc.

Assuming that the relic dark matter particles were produced soon after
symmetry breaking and that the Universe reheated at a temperature
$\trh$, we can estimate the fraction of energy in these
non-relativistic particles today as 
\be\label{sdm}
{\Omega_X h^2\over\Omega_\gamma h^2} \simeq {\rho_X\over\rho_{\rm SB}}\,
{\trh\over T_0}\,,
\ee
where $\rho_X = r\,\rho_Y$ if the relic particles come from the decay,
with branching ratio $r$, of species $Y$ produced at symmetry
breaking. Note that in the simplest case in which species $X$ couple
directly to the GUT Higgs, $r=1$, but in general we expect $r\ll1$.
Substituting the present temperature of the Universe, $T_0 = 2.73$ K
$= 2.4\times 10^{-13}$ GeV, we have
\be\label{rhoX}
{\rho_X\over\rho_{\rm SB}} = 2\times10^{-19}\,
\left({10^{10}\,{\rm GeV}\over\trh}\right)\,.
\ee
Using expressions (\ref{rhob}) and (\ref{rhof}), we see that in the
limit of small couplings, $\rho_X/\rho_{\rm SB} \simeq 2\times 10^{-3}\,
g_X^2$, where $g_X$ is the coupling to the Higgs, which gives it
a mass $M_X = g_X 10^{15}$ GeV. In order that the X-particle relics
be the cold dark matter today, we require 
\be\label{MX}
M_X = 10^7\,{\rm GeV}\,\left({10^{10}\,{\rm GeV}
\over\trh}\right)^{1/2}\,.
\ee

For a reheating temperature as high as $\trh=10^{10}$ GeV, the mass of
the particle becomes $M_X = 10^7$ GeV $< \trh$, and is thus
non-relativistic at symmetry breaking. In this case, its energy
density decays like radiation until its mass is of order the
temperature of the Universe, so we should substitute $\trh \to M_X$ in
Eqs.~(\ref{sdm}) and (\ref{rhoX}). Therefore, Eq.~(\ref{MX})
determines $M_X = 10^8$ GeV, a very natural candidate for SDM.

However, we must make sure that the SDM relics remain out of
equilibrium since their production, otherwise their thermal population
would be completely unacceptable today. For this we must ensure that
their annihilation rate is always much smaller than the expansion
rate, $\Gamma_{\rm ann} = n_X(\trh)\,\sigma_{\rm ann} |v| \ll H(\trh)$,
where the annihilation crossection is $\sigma_{\rm ann} = g_X^4/
(16\pi^2 M_X^2)$. Using expressions (\ref{nub}) and (\ref{nuf}) in the
small coupling limit, $n_X =
4.8\times10^{-4}\,\lambda\,g_X^{-2}\,M_X^3$, we get, for
$\lambda=10^{-3}$,
\be\label{eqX}
\Gamma_{\rm ann} = 3\times10^6\,{\rm GeV}\,g_X^3 \ \ll \ 
200\,{\rm GeV}\,\left({\trh\over10^{10}\,{\rm GeV}}\right)^2 = H(\trh)\,.
\ee
It is clear that for the case above, with $g_X = 10^{-7}$, we are
always safe.

Alternatively, we may consider the production of X-particles from the
decay, with very small branching ratio $r\sim10^{-10}$, of very heavy
Y-particles produced at SSB, with e.g. $M_Y = 10^{12}$ GeV and $M_X
\sim 10^{11}$ GeV. For this value of the coupling, $g_Y = 10^{-3}$, the
Y-particles remain always out of equilibrium, see (\ref{eqX}), while
their energy density at symmetry breaking is given by
$\rho_Y/\rho_{\rm SB} = 2\times10^{-9}$, giving rise to a dark matter
relic X which is compatible with present bounds.

Moreover, a very interesting consequence of this population of
extremely weakly coupled heavy relics produced at GUT symmetry
breaking is that they may constitute the origin of the ultra high
energy cosmic rays~\cite{VKT}.  In the case the X-particles are
metastable, with lifetime of the order of the age of the Universe,
their mass $M_X \sim 10^{11}$ GeV $= 10^{20}$ eV may be converted into
extremely energetic particles that reach Earth today and whose nature
could in the near future be studied by the Pierre Auger project, the
high-resolution fly's eye, and the Japanese telescope array
project~\cite{VKT}.

\section{Conclusions}

In this letter we have shown that the exponential growth of the Higgs
vev towards its true vacuum at the end of a period of hybrid
inflation, known as tachyonic preheating, can induce a significant
production of particles, both bosons and fermions coupled to the
Higgs. This new mechanism of particle production in the early Universe
could be responsible for generating the observed baryon asymmetry via
leptogenesis, as well as the present dark matter of the Universe.

We have proposed a novel scenario for leptogenesis, in which a large
population of out-of-equilibrium RH neutrinos is generated in the
process of symmetry breaking after hybrid inflation, assuming that the
symmetry breaking field is actually the Higgs field of a grand unified
theory (GUT) with ($B-L$)-violating interactions.  We have shown that
this is indeed an attractive leptogenesis scenario, since it can
explain the observed baryon asymmetry for a wide range of model
parameters.

In the case that stable relics are produced out-of-equilibrium from
the decay products of the GUT Higgs, it is possible that they may
constitute today the observed cold dark matter. These particles
may have masses as large as $10^{13}$ GeV, and could be responsible 
(if metastable) for the observed flux of ultra high energy cosmic rays.

\section*{Acknowledgements}

We are grateful to G. Felder, L. Kofman, A. Linde and I. Tkachev for
useful discussions.  This work was supported in part by the CICYT
project FPA2000-980.  J.G.B. is on leave from Universidad Aut\'onoma
de Madrid and has support from a Spanish MEC Fellowship. The work of
E.R.M. has been supported by a Marie Curie Fellowship of the European
Community TMR Program under contract HPMF-CT-2000-00581.

\newpage

\section*{Figure Captions}

{\bf Figure 1.} The time evolution of the vacuum expectation value 
  $\langle\phi^2(t)\rangle^{1/2}/v$, as compared with the approximate
  solution (\ref{phi}) with $mt_*=8$. We have used a lattice of size
  N=128 and length L=100$\pi$, which gives $k_{\rm min}=2H = 0.02m$
  and $k_{\rm max} = 2.22m$. We also show the evolution of the 
  number density, $n_{\rm B}(t)$, in units of $100\,m^{-3}$, for bosonic
  particles coupled to the Higgs with $g=0.5$.

\vspace{0.5cm}

\noindent
{\bf Figure 2.} The spectrum of occupation numbers for both bosons and 
  fermions in the (asymptotic) true vacuum, using the lattice results
  for the Higgs vev, as compared with the analytical formulae
  (\ref{nbos}) and (\ref{nfer}). The parameters chosen here are
  $\lambda=0.01$, $g=0.5$ and $h=0.5$. The tiny peaks at large momenta
  correspond to small resonances due to the strongly damped Higgs
  oscillations after SSB.

\newpage

\begin{figure}[t]
\vspace{-.5cm}
\begin{center}
\includegraphics[width=10cm,height=14cm,angle=-90]{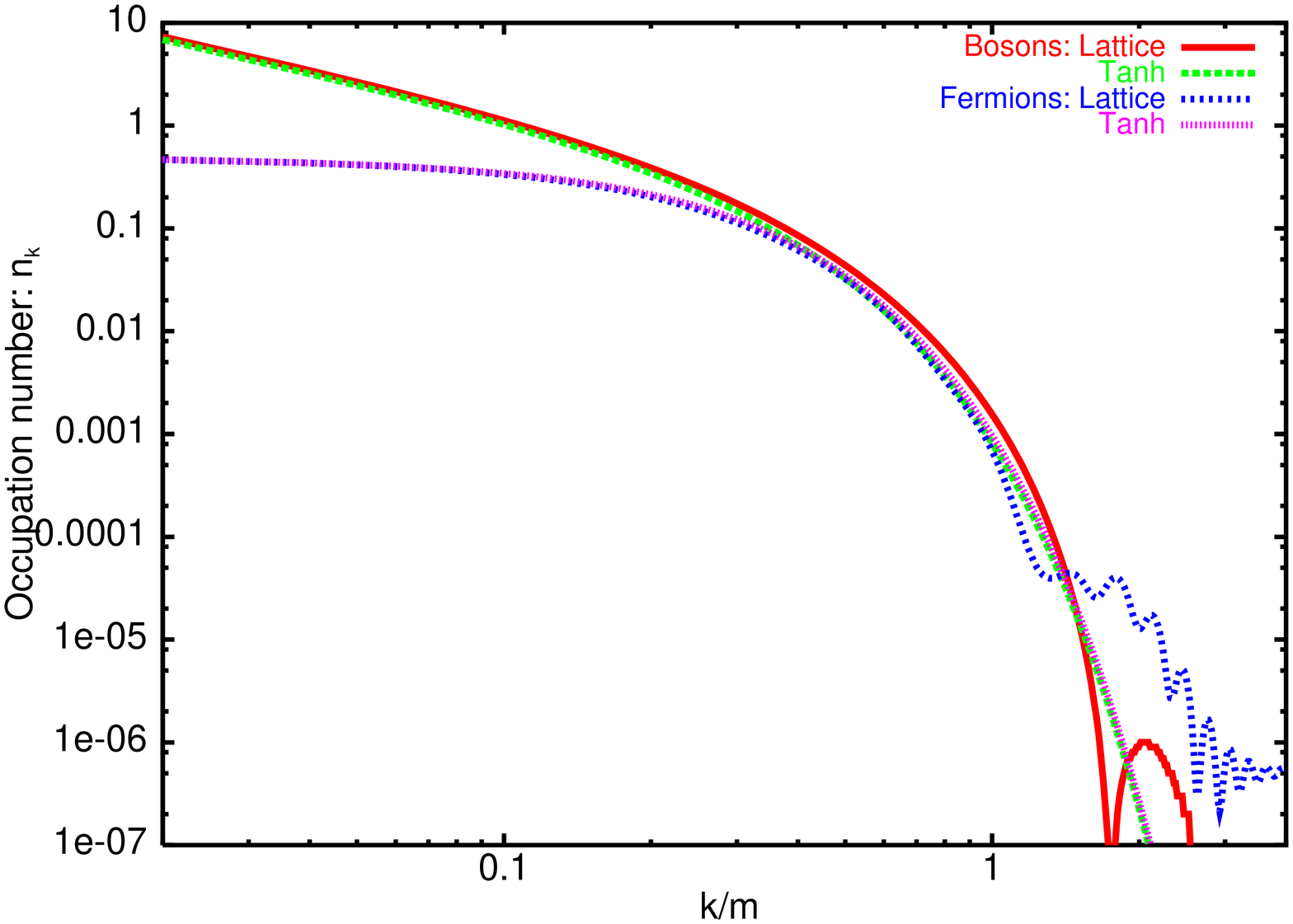}
\end{center}
\caption{} \label{fig2}
\end{figure}

\end{document}